\documentclass[prl,twocolumn,preprintnumbers,amsmath,amssymb,superscriptaddress]{revtex4-2}

\usepackage{amsmath}
\usepackage{tabularx}
\usepackage{graphicx}
\usepackage{dcolumn}
\usepackage{bm}
\usepackage{hyperref}
\begin{document}
\title{Singular viscoelastic perturbation to soft lubrication}
\author{Bharti Bharti}
\thanks{These two authors contributed equally.}
\affiliation{Univ. Bordeaux, CNRS, LOMA, UMR 5798, F-33400, Talence, France.}
\affiliation{Mechanics Division, Department of Mathematics, University of Oslo, Oslo 0316, Norway.}
\thanks{These two authors contributed equally.}
\author{Quentin Ferreira}
\thanks{These two authors contributed equally.}
\affiliation{Univ. Bordeaux, CNRS, LOMA, UMR 5798, F-33400, Talence, France.}
\author{Aditya Jha}
\affiliation{Univ. Bordeaux, CNRS, LOMA, UMR 5798, F-33400, Talence, France.}
\author{Andreas Carlson}
\affiliation{Mechanics Division, Department of Mathematics, University of Oslo, Oslo 0316, Norway.}
\author{David Dean} 
\affiliation{Univ. Bordeaux, CNRS, LOMA, UMR 5798, F-33400, Talence, France.}
\author{Yacine Amarouchene}
\affiliation{Univ. Bordeaux, CNRS, LOMA, UMR 5798, F-33400, Talence, France.}
\author{Tak Shing Chan}
\affiliation{Mechanics Division, Department of Mathematics, University of Oslo, Oslo 0316, Norway.}
\author{Thomas Salez}\email{thomas.salez@cnrs.fr}
\affiliation{Univ. Bordeaux, CNRS, LOMA, UMR 5798, F-33400, Talence, France.}
\date{\today}
\begin{abstract}
Soft lubrication has been shown to drastically affect the mobility of an object immersed in a viscous fluid in the vicinity of a purely elastic wall. In this theoretical study, we develop a minimal model incorporating  viscoelasticity, carrying out a  perturbation analysis in both the elastic deformation of the wall and its viscous damping. Our approach reveals the singular-perturbation nature of viscoelasticity to soft lubrication. Numerical resolution of the resulting non-linear, singular and coupled equations of motion reveals  peculiar effects of viscoelasticity on confined colloidal mobility, opening the way towards the description of complex migration scenarios near realistic polymeric substrates and biological membranes.
\end{abstract}                   
\maketitle

The dynamics of objects moving in lubricated environnements near rigid boundaries has been extensively studied over the last century~\cite{goldman1967slow,o1967slow,cooley1969slow,jeffrey1981slow}, with the aim of characterizing interfacial colloidal transport. Experimental investigations and theoretical studies have shown that the associated modification of the boundary conditions leads to anisotropic and space-dependent mobilities.
As a consequence, Brownian particles exhibit non-Gaussian features~\cite{Felderhof2005, Elgeti2015, Jeney2008, Huang2015, Choudhury2017, Hertlein2008, Helden2015, Matse2017, Alexandre2023}. The introduction  of elasticity in the objects interacting with the fluid flow underpins the behaviour in settings ranging from roller bearings to the sliding friction of snow, and even the motion of blood cells through capillaries, to name but a few~\cite{Dowson&Hgginson1959, Archard1961, Crook1961, Glenne1987, Campbell1989, Lighthill1968, Gerald1969, Chan1979, Ma2003}.

The elastic response of a soft boundary, and its consequences, due to rotation and translation of a neighbouring object in lubricated contact has been extensively studied~\cite{sekimoto1993mechanism, Beaucourt2004,Skotheim2004b, Skotheim2005, Urzay2007, Urzay2010, Leroy2011,Snoeijer2013, Saintyves2016,Rallabandi2017, Dillard2018,Saintyves2020}. When submerged in a fluid near a deformable solid, the friction force acting on a moving object is described through an elastohydrodynamic (EHD) coupling within the lubrication approximation. Experiments based on Surface Forces Apparatus (SFA) and Atomic Force Microscopy (AFM) have not only detected the EHD coupling~\cite{Leroy2012,Vialar2019,Zhang2020}, but have also been used to extract the mechanical properties of the solid or liquid deformable surfaces~\cite{Guan2017, Wang2017a, Wang2017b}. Oscillating colloidal probes in lubricated contacts have become standard, powerful, non-invasive and contactless probes of the rheological properties of materials~\cite{garcia2016micro, basoli2018biomechanical}. Further investigations of electrokinetic effects~\cite{Wong2003, Chakraborty2010}, and non-linear responses of the solid boundary~\cite{Davies2018,Rallabandi2018,Daddi2018}, with the goal of understanding EHD in the context of biological membranes, has also been performed. Finally, the  resulting coupling between several degrees of freedom can change the object's dynamics in surprising ways, as illustrated by theoretical predictions in the purely elastic case~\cite{Salez2015,Bertin2022}, where a cylinder or a sphere can simultaneously sediment, slide and spin near a soft wall. 

While most soft materials also exhibit viscoelasticity, the inclusion of the latter on the motion of a nearby object has only been analysed with a restricted number of degrees of freedom~\cite{Pandey2016, Maali2017,Snoeijer2020, Estahbanati2021,Haim2022, Hu2023}. In the present work, we  aim to uncover and characterize the key effects of viscoelasticity on soft lubrication. To do so, we extend the purely elastic Winkler soft-lubrication framework~\cite{Dillard2018,Chandler2020} of the free-particle problem~\cite{Salez2015}, to the viscoelastic case. Specifically, we invoke a Kelvin-Voigt response as a perturbation to the Winkler model. The resulting modified soft-lubricated dynamics of a moving cylinder is then  investigated by calculating the force corrections at leading order in viscoelastic compliance, and exhibits an unexpected singular-perturbation nature, in addition to  original effects with respect to the purely elastic case.

\begin{figure}[!thbp]
\begin{center}
\includegraphics[width=0.60\linewidth]{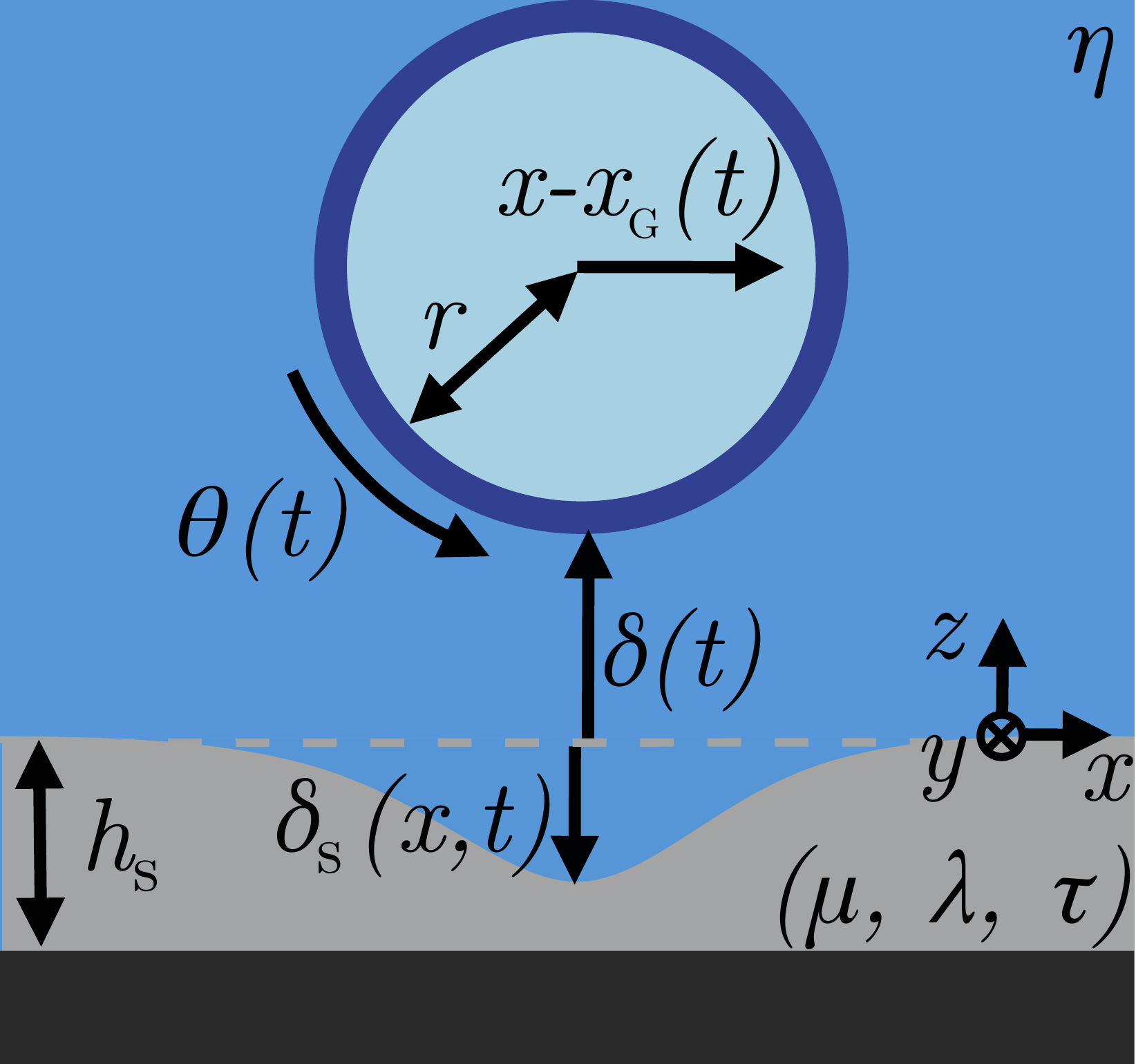}
\caption{Schematic representation of a moving cylinder of radius $r$ at time $t$ in a viscous fluid of dynamic shear viscosity $\eta$, near and above a thin viscoelastic compressible wall of thickness $h_{\textrm{s}}$, shear elastic modulus $\mu$, Lam\'e coefficient $\lambda$, and Kelvin-Voigt viscoelastic response time $\tau$. The degrees of freedom of the cylinder are the horizontal, vertical and angular positions, denoted by $x_{\textrm{G}}(t)$, $\delta(t)$, and $\theta(t)$, respectively. The wall deformation profile is denoted by $\delta_{\textrm{s}}(x,t)$. }
\label{fig:1}
\end{center}
\end{figure}
We consider a freely-moving rigid infinite cylinder in a viscous fluid and near a thin compressible viscoelastic wall, as described in details in Fig.~\ref{fig:1}. Fluid inertia is neglected. We assume the lubrication approximation to be valid, \textit{i.e.} that the typical distance $\delta_0$ between the cylinder and the undeformed substrate satisfies $\delta_0=\epsilon r$ with $\epsilon\ll 1$. The horizontal fluid velocity field is thus dominant and is denoted by $u(x,z,t)$. The hydrodynamic pressure field $p(x,t)$, in excess to the atmospheric pressure, is invariant along $z$ and assumed to vanish as $x\rightarrow \pm \infty$. In the limit of small gap, the profile of the cylinder can be approximated by a parabola, leading to the total gap profile:
\begin{equation}\label{1}
h(x,t)\simeq\delta (t)-\delta_{\textrm{s}}(x,t)+\frac{\left[x-x_{\textrm{G}}(t)\right]^2}{2r}\,.
\end{equation}
The substrate deformation is described through a combination of a Winkler foundation and a Kelvin-Voigt-like viscoelastic perturbation, as: 
\begin{equation}\label{2}
\delta_{\textrm{s}}(x,t)\simeq-\frac{h_{\textrm{s}}}{2\mu+\lambda}\left[p(x,t)-\tau\dot{p}(x,t)\right]\,,
\end{equation}
where the dot corresponds to a time derivative, and $\tau$ denotes the viscoelastic response time of the substrate -- assumed to be vanishingly small compared to the time scale of motion.

Let us now non-dimensionalize the problem through: $z=Zr\epsilon$, $h=Hr\epsilon$, $\delta=\Delta r\epsilon$, $x=X r\sqrt{2\epsilon}$, $x_{\textrm{G}}=X_{\textrm{G}}r\sqrt{2\epsilon}$, $\theta=\Theta\sqrt{2\epsilon}$, $t=Tr\sqrt{2\epsilon}/c$, $u=Uc$, $p=P\eta c \sqrt{2}/(r\epsilon^{3/2})$, where $c$ is a horizontal velocity scale. The dimensionless total gap profile thus reads:
\begin{equation}
\small
H(X,T)=\Delta(T)+ \left[X-X_{\textrm{G}}(T)\right]^2+\kappa P(X,T)-\beta \dot{P}(X,T)\,,
\end{equation}
where we have introduced the elastic compliance $\kappa=\sqrt{2}h_{\textrm{s}}\eta c/[r^2\epsilon^{5/2}(2\mu+\lambda)]$,
and the viscoelastic compliance $\beta=h_{\textrm{s}}\eta\tau c^2/[r^{3}\epsilon^{3}(2\mu+\lambda)]$.

The fluid flow is governed by the incompressible steady Stokes equation, which in the lubrication approximation is:
\begin{equation}\label{6}
 U_{ZZ}=P_X\,, \hspace{10pt} P_Z = 0\,,
\end{equation}
where the indices denote spatial derivatives. We assume no-slip boundary conditions at both the cylinder and substrate surfaces, \textit{i.e.} $U(X,Z=-\kappa P+\beta \dot{P},T)=0$ and  $U(X,Z=H-\kappa P+\beta \dot{P},T)=\dot{X}_{\textrm{G}}+\dot{\Theta}$. Furthermore, volume conservation implies:
\begin{equation} \label{vol}
\partial_TH+\partial_X\int_{-\kappa P+\beta\dot{P}}^{H-\kappa P+\beta\dot{P}} \textrm{d}Z\, U=0 \,.
\end{equation}
Solving the Stokes equations for the velocity field, and injecting the solution into Eq.~(\ref{vol}) then leads to the Reynolds equation:
\begin{equation}\label{8}
\begin{split}
12\dot{\Delta}-24(X-X_{\textrm{G}})&\dot{X}_{\textrm{G}}+12\kappa \dot{P}-12\beta \ddot{P}\\
&=\left[H^3P_X-6(\dot{X}_{\textrm{G}}+\dot{\Theta})H\right]_X\,.
\end{split}
\end{equation}

We now define the dimensionless flow-induced forces exerted on the cylinder. At vanishing $\epsilon$, the pressure-induced drag force per unit length along $Z$ can be approximated by:
\begin{equation}
D_{p,\perp}\simeq\int_{- \infty}^{ \infty}\textrm{d}X\, P\,, \label{2}
\end{equation}
while the pressure-induced drag force per unit length along $X$ can be approximated by:
\begin{equation}
D_{p,\parallel}\simeq-\sqrt{2\epsilon}\int_{- \infty}^{ \infty}\textrm{d}X\, (X-X_{\textrm{G}})P\,.\label{3}
\end{equation}
Finally, the shear-induced drag force per unit length along $X$ can be approximated by:
\begin{equation}
D_{\sigma,\parallel }\simeq-\sqrt{\frac{\epsilon}{2}}\int_{- \infty}^{ \infty}\textrm{d}X\, \left.U_Z\right|_{Z=H-\kappa P+\beta\dot{P}}\,.\label{4}
\end{equation}
Note that the corresponding dimensional forces per unit length $d_{m,n}$, with $m=p,\sigma$ and $n=\perp,\parallel$, can be obtained from the definition $d_{m,n}=2c\eta D_{m,n}/\epsilon$.

Following previous studies on soft lubrication~\cite{Skotheim2005,Urzay2007,Leroy2011,Salez2015,Wang2015,Rallabandi2017,Bertin2022}, we assume that the deformation is small compared to the other typical length scales of the problem, implying that $\kappa\ll1$ and $\beta\ll1$. Since our primary goal is to address the effect of the viscoelastic perturbation with respect to the purely elastic case, we further introduce the definition $\beta=\kappa\alpha$ and assume $\alpha\ll1$. Then, we invoke a perturbation expansion of the pressure field, at first order in $\kappa$ and $\alpha$, by writing:
\begin{equation}\label{12}
P\simeq P^{(00)}+\kappa \left[P^{(10)}-\alpha P^{(11)}\right]\,,
\end{equation}
where $P^{(ij)}$ characterizes the magnitude of the pressure correction at orders $i$ in $\kappa$ and $j$ in $\alpha$. Details on the derivation of $P^{(ij)}$, and the associated magnitudes of the flow-induced dimensionless forces per unit length $D^{(i,j)}_{m,n}$, with $m=p,\sigma$ and $n=\perp,\parallel$, are provided in the Supplementary Material (SM).  

Using the resulting flow-induced forces per unit length, conservation of linear and angular momenta eventually leads to the following three equations of motion of the free cylinder:
\small
\begin{widetext}
\begin{equation}
\begin{split}
&  \ddot{\Delta}+\xi\frac{\dot{\Delta}}{\Delta^{3/2}}+\frac{\kappa\xi}{4}\left[21\frac{\dot{\Delta}^2}{\Delta^{9/2}}-\frac{\left(\dot{\Theta}-\dot{X}_{\textrm{G}}\right)^2}{\Delta^{7/2}}-\frac{15}{2}\frac{\ddot{\Delta}}{\Delta^{7/2}}\right]-\frac{\beta\xi}{4}\left[\frac{189}{4}\frac{\dot{\Delta}}{\Delta^{9/2}}\left(\ddot{\Delta}-\frac{\dot{\Delta}^2}{\Delta}\right)-\frac{\ddot{\Theta}}{2\Delta^{7/2}}\left(2\dot{\Theta}-7\dot{X}_{\textrm{G}}\right)\right.\\
&\left.-\frac{\ddot{X}_{\textrm{G}}}{2\Delta^{7/2}}\left(12\dot{X}_{\textrm{G}}-7\dot{\Theta}\right)+\frac{7}{4}\frac{\dot{\Delta}}{\Delta^{9/2}}\left(6\dot{X}_{\textrm{G}}^2+\dot{\Theta}^2-7\dot{X}_{\textrm{G}}\dot{\Theta}\right)-\frac{15}{2}\frac{\dddot{\Delta}}{\Delta^{7/2}}\right]=0~,\label{5} 
\end{split}
\end{equation}
\begin{equation}
\begin{split}
& \ddot{X}_{\textrm{G}}+\frac{2\epsilon\xi}{3}\frac{\dot{X}_{\textrm{G}}}{\Delta^{1/2}}+\frac{\kappa\epsilon\xi}{6}\left(\frac{19}{4}\frac{\dot{\Delta}\dot{X}_{\textrm{G}}}{\Delta^{7/2}}-\frac{\dot{\Delta}\dot{\Theta}}{\Delta^{7/2}}+\frac{1}{2}\frac{\ddot{\Theta}-\ddot{X}_{\textrm{G}}}{\Delta^{5/2}}\right)-\frac{\beta\epsilon\xi}{6}\left[\frac{21}{16}\frac{\dot{\Delta}^2}{\Delta^{9/2}}\left(\dot{\Theta}-9\dot{X_{\textrm{G}}}\right)+\frac{9}{8}\frac{\dot{X}_{\textrm{G}}}{\Delta^{7/2}}\left(\dot{\Theta}-\dot{X}_{\textrm{G}}\right)^2 \right.\\
& \left.+\frac{\ddot{\Delta}}{2\Delta^{7/2}}\left(\frac{1}{4}\dot{\Theta}+11\dot{X}_{\textrm{G}}\right)+\frac{\dot{\Delta}}{2\Delta^{7/2}}\left(\frac{17}{2}\ddot{X}_{\textrm{G}}-\frac{19}{4}\ddot{\Theta}\right)+\frac{\dddot{\Theta}-\dddot{X}_{\textrm{G}}}{2\Delta^{5/2}}\right]=0~,\label{6} 
\end{split}
\end{equation}
\begin{equation}
\begin{split}
&  \ddot{\Theta}+\frac{4\epsilon\xi}{3}\frac{\dot{\Theta}}{\Delta^{1/2}}+\frac{\kappa\epsilon\xi}{3}\left(\frac{19}{4}\frac{\dot{\Delta}\dot{\Theta}}{\Delta^{7/2}}-\frac{\dot{\Delta}\dot{X}_{\textrm{G}}}{\Delta^{7/2}}+\frac{1}{2}\frac{\ddot{X}_{\textrm{G}}-\ddot{\Theta}}{\Delta^{5/2}}\right)-\frac{\beta\epsilon\xi}{3}\left[\frac{21}{16}\frac{\dot{\Delta}^2}{\Delta^{9/2}}\left(5\dot{X}_{\textrm{G}}-7\dot{\Theta}\right)-\frac{9}{8}\frac{\dot{X}_{\textrm{G}}}{\Delta^{7/2}}\left(\dot{\Theta}-\dot{X}_{\textrm{G}}\right)^2 \right.\\
&\left.+\frac{\ddot{\Delta}}{2\Delta^{7/2}}\left(11\dot{\Theta}-\frac{29}{4}\dot{X}_{\textrm{G}}\right)+\frac{\dot{\Delta}}{2\Delta^{7/2}}\left(\ddot{\Theta}-\frac{19}{4}\ddot{X}_{\textrm{G}}\right) +\frac{\dddot{X}_{\textrm{G}}-\dddot{\Theta}}{2\Delta^{5/2}}\right]=0\,~,\label{7} 
\end{split}
\end{equation}
\end{widetext}
where we have further introduced the dimensionless dissipation parameter $\xi=6\eta/(\epsilon r\rho c)$, with $\rho$ the mass density of the cylinder. As in the purely-elastic case, the governing equations are coupled, non-linear, singular, as well as inertial-like, and they involve a large number of parameters and variables~\cite{Salez2015,Bertin2022}. However, viscoelasticity breaks the symmetry between the sliding and rotational degrees of freedom observed in the purely-elastic case. Moreover, strikingly, novel jerk-like terms -- often associated with dynamical systems and chaotic behavior~\cite{Eichhorn2002,Sprott2009} and appearing \textit{e.g.} in the theory of radiating electrons~\cite{Dirac1938} -- arise due to the addition of viscoelasticity in the substrate. Equations~(\ref{5}),~(\ref{6}),~and (\ref{7}) can be numerically integrated (see SM) to obtain the trajectory of the cylinder for given initial conditions. In the following, we address two canonical start-up problems which illustrate the new effects induced by viscoelasticity. 

\begin{figure}[!h]
\centering
\includegraphics[width=\linewidth]{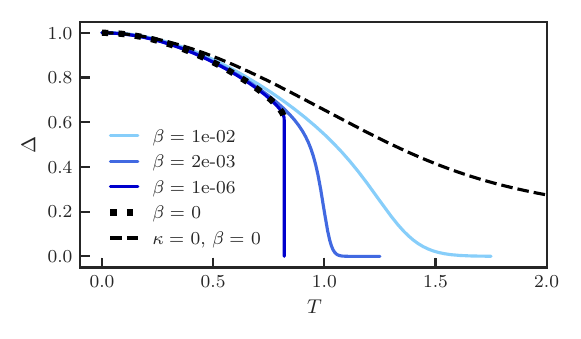}
\caption{Dimensionless distance $\Delta$ between the cylinder and the undeformed wall as a function of dimensionless time $T$, as obtained from numerically solving Eqs.~(\ref{5}),~(\ref{6}),~and (\ref{7}), with the addition of a -1 constant term on the right-hand side of Eq.~(\ref{5}). The black dashed line corresponds to a rigid wall ($\kappa=0$,$\beta=0$). The black dotted line corresponds to a purely elastic substrate ($\beta=0$) with $\kappa=0.1$. The coloured solid lines correspond to viscoelastic substrates with $\kappa=0.1$ and three values of $\beta$, as indicated. All variables and their time derivatives are initialized to zero, except for $\Delta(T=0)=1$. The common fixed parameters are $\xi=1$ and $\epsilon=0.1$.}
\label{fig:sedim}
\end{figure}

First, we consider a sedimentation scenario, where gravity appears as  a constant  negative  term on the right-hand side of Eq.~(\ref{5}). The numerical solutions are shown in Fig.~\ref{fig:sedim}. In the purely-elastic case ($\beta=0$), due to the EHD coupling and the absence in the model of any fluid-inertial regularization mechanism at short times, the acceleration of the cylinder eventually diverges. This divergence stems from  the vanishing of the effective inertial mass~\cite{Salez2015}, which occurs at a critical vertical coordinate $\Delta_\mathrm{c}=(15\kappa\xi/8)^{2/7}\approx0.62$, in agreement with the numerical solution. However, when $\beta\neq0$, the divergence is no longer observed due to viscoelastic regularization. Indeed, the effective inertial mass becomes $1-15\kappa\xi/(8\Delta^{7/2})-189\beta\xi\dot\Delta/(16\Delta^{9/2})$, which remains finite in general. In addition, increasing $\beta$ leads to an increase of the sedimentation time. The latter observation agrees with intuition, as large viscoelastic times tend to delay -- and hence effectively reduce -- the substrate's elastic response, which thus becomes closer to that of a rigid wall.

\begin{figure}[!h]
\centering
\includegraphics[width=\linewidth]{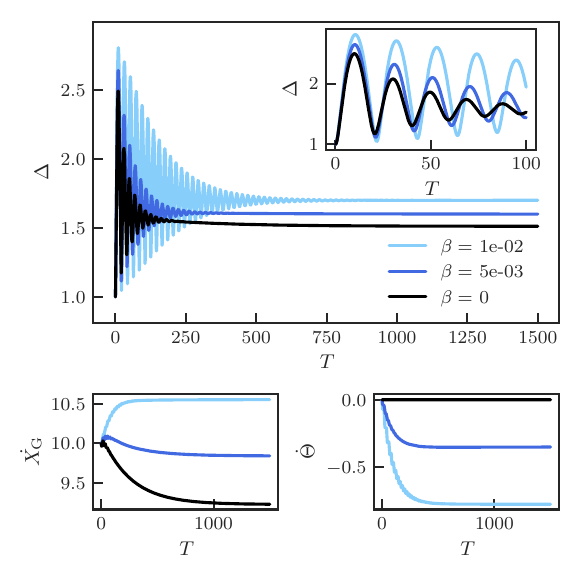}
\caption{Dimensionless distance $\Delta$ between the cylinder and the undeformed wall, dimensionless translational speed $\dot{X}_{\textrm{G}}$ along the wall, and dimensionless angular speed $\dot{\Theta}$, as functions of dimensionless time $T$, as obtained from numerically solving Eqs.~(\ref{5}),~(\ref{6}),~and (\ref{7}), with the additions of $-0.05$ and $0.05$ constant terms on the right-hand sides of Eqs.~(\ref{5}) and~Eq.~(\ref{6}), respectively. The black solid line corresponds to a purely elastic substrate ($\beta=0$). The coloured solid lines correspond to viscoelastic substrates with two values of $\beta$, as indicated. All variables and their time derivatives are initialized to zero, except for $\Delta(T=0)=1$ and $\dot{X}_{\textrm{G}}(T=0)=10$. The common fixed parameters are $\kappa=0.1$, $\xi=1$ and $\epsilon=0.1$. The inset is a zoom at short times.}
\label{fig:start-up}
\end{figure}

One of the most striking effects of soft lubrication is the emergence of a lift force at zero Reynolds number~\cite{Bureau2023,Rallabandi2024}. Consequently, the second  scenario we consider is  lubricated sliding along an inclined wall~\cite{Saintyves2016}. To do so, we add constant gravity-like components in the right-hand sides of Eqs.~(\ref{5}) and~(\ref{6}), as well as a non-zero initial transverse velocity $\dot{X}_{\textrm{G}}(T=0)$ in order to allow for immediate take off. The numerical solutions are shown in Fig.~\ref{fig:start-up}. We observe damped oscillations of the normal position $\Delta$ toward a long-term steady state, the amplitude, frequency, and damping of which appear to be affected by viscoelasticity. 

The most striking and counterintuitive result on the latter sliding scenario is perhaps the increase of the steady altitude with $\beta$. Indeed, previous theoretical studies~\cite{Pandey2016,Estahbanati2021,Hu2023} have predicted the opposite, \textit{i.e.} that viscoelasticity inhibits lift. However, these results were obtained with a restricted number of degrees of freedom, namely at constant distance from the wall, at constant translational velocity along the wall and 
without allowing for rotation. In the latter conditions, our model would react (not shown) exactly as expected for our specific viscoelastic constitutive law, \textit{i.e.} the steady value of $\Delta$, and hence the lift force, would be independent of $\beta$. All together, this reveals that the interplay of the multiple degrees of freedom of the cylinder is crucial to understand the full impact of wall viscoelasticity on the hydrodynamic mobility of a neighbouring object. To corroborate this statement, Fig.~\ref{fig:start-up} also reveals an increase in steady translational velocity $\dot{X}_{\textrm{G}}$ along the wall, as well as the apparition of a non-zero and significant steady angular velocity $\dot{\Theta}$, as a result of viscoelasticity. The difference between these two velocities -- which increases in magnitude with $\beta$ -- greatly influences the lift, as can be predicted from the steady state (denoted by the $\infty$ subscript) of Eq.~(\ref{5}): $\Delta_\mathrm{\infty}=[\kappa\xi(\dot{\Theta}_{\infty}-\dot{X}_{G \infty})^2/(4F)]^{2/7}$, where $F$ is the positive magnitude of the added negative gravity-like constant force. 

To conclude, the impact of boundary viscoelasticity on the hydrodynamic mobility of a nearby object proves to be more complex and interesting than previously anticipated. Indeed, it does not only boil down to additional damping and temporal regularization of sharp elastic behaviours. Principally, viscoelasticity appears as a singular perturbation to purely-elastic soft lubrication, as it generates higher-order jerk-like differential terms usually associated with the rich physics of dynamical systems, and as it breaks the symmetry between the sliding and rotational degrees of freedom observed in the purely-elastic case. Moreover, when  all  degrees of freedom of the moving object are taken into account, it generates non-trivial effects such as the enhancement of the soft-lubrication lift force -- in sharp contrast with previous predictions on more-constrained motions. In addition, while the current calculations have been performed in a two-dimensional setting, we expect the qualitative properties encountered here to hold in three dimensions, by analogy with previous works~\cite{Salez2015,Bertin2022}. As such, these key modifications, introduced into the soft-lubrication picture by viscoelasticity, could enable a realistic comparison with AFM/SFA contactless colloidal-probe experiments near  elastomers, gels, lipid bilayers or fluid interfaces. Furthermore, applying the fluctuation-dissipation theorem to the present results, Brownian motion is expected to be directly and non-trivially affected by the vicinity of viscoelastic boundaries. We thus anticipate important implications in nanoscience and biophysics, where the transport of microscopic entities near complex interfaces is ubiquitous. 

\begin{acknowledgments}
The authors thank Pierre Gresil, Yilin Ye and Haim Diamant for interesting discussions. They acknowledge financial support from the European Union through the European Research Council under EMetBrown (ERC-CoG-101039103) grant, from the Agence Nationale de la Recherche under the Softer (ANR21-CE06-0029) and Fricolas (ANR-21-CE06-0039) grants, and from the Research Council of Norway (Project No. 315110). 
\end{acknowledgments}
\bibliography{Bharti2024}
\end{document}


\title{Singular viscoelastic perturbation to soft lubrication\\-- Supplementary Material --}
\author{Bharti Bharti}
\thanks{These two authors contributed equally.}
\affiliation{Univ. Bordeaux, CNRS, LOMA, UMR 5798, F-33400, Talence, France.}
\affiliation{Mechanics Division, Department of Mathematics, University of Oslo, Oslo 0316, Norway.}
\thanks{These two authors contributed equally.}
\author{Quentin Ferreira}
\thanks{These two authors contributed equally.}
\affiliation{Univ. Bordeaux, CNRS, LOMA, UMR 5798, F-33400, Talence, France.}
\author{Aditya Jha}
\affiliation{Univ. Bordeaux, CNRS, LOMA, UMR 5798, F-33400, Talence, France.}
\author{Andreas Carlson}
\affiliation{Mechanics Division, Department of Mathematics, University of Oslo, Oslo 0316, Norway.}
\author{David Dean} 
\affiliation{Univ. Bordeaux, CNRS, LOMA, UMR 5798, F-33400, Talence, France.}
\affiliation{Team MONC, INRIA Bordeaux Sud Ouest, CNRS UMR 5251, Bordeaux INP, Univ. Bordeaux, F-33400 Talence, France.}
\author{Yacine Amarouchene}
\affiliation{Univ. Bordeaux, CNRS, LOMA, UMR 5798, F-33400, Talence, France.}
\author{Tak Shing Chan}
\affiliation{Mechanics Division, Department of Mathematics, University of Oslo, Oslo 0316, Norway.}
\author{Thomas Salez}\email{thomas.salez@cnrs.fr}
\affiliation{Univ. Bordeaux, CNRS, LOMA, UMR 5798, F-33400, Talence, France.}
\date{\today}             
\maketitle
\section{Rigid case}\label{appa}
Notations are all defined in the main text. In the absence of wall deformation, the fluid flow is governed by the incompressible Stokes equation in the lubrication approximation, which reads: 
\begin{equation}\label{A1}
U_{ZZ}=P_X\ .
\end{equation}
The fluid gap profile can be approximated by a parabola, as:
\begin{equation}\label{A2}
H(X,T)\simeq\Delta (T)+ [X-X_{\textrm{G}}(T)]^2\ .
\end{equation}
Invoking no-slip boundary conditions, $U(X,Z=0,T)=0$ and $U(X,Z=H,T)=\dot{X}_{\textrm{G}}+\dot{\Theta}$, the Stokes equation can be solved, leading to the horizontal velocity profile:
\begin{equation}\label{A3}
U=\frac{P_X}{2}Z\left[Z-\Delta -(X-X_{\textrm{G}})^2\right]+\frac{(\dot{X}_{\textrm{G}}+\dot{\Theta})Z}{\Delta + \left(X-X_{\textrm{G}}\right)^2}\ .
\end{equation}
Volume conservation reads:
\begin{equation}
\partial_TH+\partial_X\int_{0}^{H}\textrm{d}Z\, U=0\ .
\end{equation}
Integrating the latter, and assuming a vanishing lubrication pressure $P$ at $X\rightarrow\pm \infty$, gives: 
\begin{equation}\label{A4}
P=-\frac{3\dot{\Delta}+2(\dot{\Theta}-\dot{X}_{\textrm{G}})(X-X_{\textrm{G}})}{\left[\Delta + \left(X-X_{\textrm{G}}\right)^2\right]^2}\ .
\end{equation}
The pressure-induced drag force (per unit length) along $Z$ is given by:
\begin{equation}\label{A5}
D_{p,\perp}=\int_{- \infty}^{ \infty}\textrm{d}X\, P=-\frac{3\pi}{2}\frac{\dot{\Delta}}{\Delta^{3/2}}\ .
\end{equation}
The pressure-induced drag force (per unit length) along $X$ is given by:
\begin{equation}\label{A6}
D_{p,\parallel}=-\sqrt{2\epsilon}\int_{- \infty}^{ \infty}\textrm{d}X\, (X-X_{\textrm{G}})P=\pi \sqrt{2\epsilon}\frac{\dot{\Theta}-\dot{X}_{\textrm{G}}}{\Delta^{1/2}}\ .
\end{equation}
The shear-induced drag force (per unit length) along $X$ is given by:
\begin{equation}\label{A7}
D_{\sigma,\parallel }=-\sqrt{\frac{\epsilon}{2}}\int_{- \infty}^{ \infty}\textrm{d}X\, U_Z|_{Z=H}=-\pi\sqrt{2\epsilon}\frac{\dot{\Theta}}{\Delta^{1/2}}\ .
\end{equation}

\section{Elastic correction}\label{appb}
The fluid gap profile becomes:
\begin{equation}
H(X,T)\simeq\Delta (T)+ [X-X_{\textrm{G}}(T)]^2+\kappa P(X,T)\ .
\end{equation}
The new no-slip boundary conditions are: $U(X,Z=-\kappa P,T)=0$  and $U(X,Z=H-\kappa P,T)=\dot{X}_{\textrm{G}}+\dot{\Theta}$. Volume conservation now reads:
\begin{equation}\label{B1}
\partial_TH+\partial_X\int_{-\kappa P}^{H-\kappa P}\textrm{d}Z\, U=0\ ,
\end{equation}
which leads to the Reynolds equation:
\begin{equation}\label{B2}
12\dot{\Delta}-24(X-X_{\textrm{G}})\dot{X}_{\textrm{G}}+12\kappa \dot{P}=\left[H^3P_X-6(\dot{X}_{\textrm{G}}+\dot{\Theta})H\right]_X\ .
\end{equation}
We now invoke a perturbation analysis at first order in $\kappa$, as: 
\begin{equation}\label{B3}
\begin{split}
&P\simeq P^{(00)}+\kappa P^{(10)}\ ,\\
&D_{p, \perp}\simeq D_{p, \perp}^{(00)}+\kappa D_{p, \perp}^{(10)}\ ,\\
&D_{p, \parallel}\simeq D_{p, \parallel}^{(00)}+\kappa D_{p, \parallel}^{(10)}\ ,\\
&D_{\sigma, \parallel}\simeq D_{\sigma, \parallel}^{(00)}+\kappa D_{\sigma, \parallel}^{(10)}\ .
\end{split}
\end{equation}
The first-order correction to the Reynolds equation then reads:
\begin{equation}\label{B4}
\begin{split}
& \left\{\left[\Delta+(X-X_{\textrm{G}})^2\right]^3P_X^{(10)}+3\left[\Delta+\left(X-X_{\textrm{G}}\right)^2\right]^2P^{(00)}P_X^{(00)}\right.\\
&\left. -6\left(\dot{X}_{\textrm{G}} +\dot{\Theta}\right)P^{(00)}\right\}_X=12P_T^{(00)}\ .
\end{split}
\end{equation}
From the previous section, one has:
\begin{equation}
\begin{split}
&P^{(00)}=-\frac{3\dot{\Delta}+2(\dot{\Theta}-\dot{X}_{\textrm{G}})(X-X_{\textrm{G}})}{\left[\Delta + \left[X-X_{\textrm{G}}\right]^2\right]^2}\ ,\\
&D_{p, \perp}^{(00)}=-\frac{3\pi}{2}\frac{\dot{\Delta}}{\Delta^{3/2}}\ ,\\
&D_{p, \parallel}^{(00)}=\pi \sqrt{2\epsilon}\frac{\dot{\Theta}-\dot{X}_{\textrm{G}}}{\Delta^{1/2}}\ ,\\
&D_{\sigma, \parallel}^{(00)}=-\pi\sqrt{2\epsilon}\frac{\dot{\Theta}}{\Delta^{1/2}}\ .
\end{split}    
\end{equation}\\
Solving Eq.~(\ref{B4}) with vanishing $P^{(10)}$ at $X\rightarrow\pm \infty$ leads to:
\begin{equation}
\begin{split}
&D_{p, \perp}^{(10)}=\frac{45\pi \ddot{\Delta}}{16 \Delta^{7/2}}-\frac{63\pi \dot{\Delta}^2}{8 \Delta^{9/2}}+\frac{3\pi (\dot{\Theta}-\dot{X}_{\textrm{G}})^2}{8 \Delta^{7/2}}\ ,\\
&D_{p, \parallel}^{(10)}=\sqrt{\frac{\epsilon}{2}}\left[\frac{23\pi \dot{\Delta}(\dot{\Theta}-\dot{X}_{\textrm{G}})}{8 \Delta^{7/2}}+\frac{\pi(\ddot{X}_{\textrm{G}}-\ddot{\Theta})}{2 \Delta^{5/2}}\right]\ ,\\
&D_{\sigma, \parallel}^{(10)}=\sqrt{\frac{\epsilon}{2}}\left[\frac{\pi (\ddot{\Theta}-\ddot{X}_{\textrm{G}})}{4 \Delta^{5/2}}+\frac{\pi \dot{\Delta} \dot{X_{\textrm{G}}}}{2\Delta^{7/2}}-\frac{19\pi \dot{\Delta} \dot{\Theta}}{8\Delta^{7/2}}\right]\ .
\end{split}
\end{equation}

\section{Viscoelastic correction}\label{appc}
The fluid gap profile becomes:
\begin{equation}
H(X,T)\simeq\Delta (T)+ [X-X_{\textrm{G}}(T)]^2+\kappa P(X,T)-\beta\dot{P}(X,T)\ .
\end{equation}
The new no-slip boundary conditions are: $U(X,Z=-\kappa P+\beta\dot{P},T)=0$  and $U(X,Z=H-\kappa P+\beta\dot{P},T)=\dot{X}_{\textrm{G}}+\dot{\Theta}$. Volume conservation now reads:
\begin{equation}\label{C1}
\partial_TH+\partial_X\int_{-\kappa P+\beta\dot{P}}^{H-\kappa P+\beta\dot{P}}\textrm{d}Z\, U=0\ ,
\end{equation}
which leads to the Reynolds equation:
\begin{equation}\label{C2}
\small
12\dot{\Delta}-24(X-X_{\textrm{G}})\dot{X}_{\textrm{G}}+12\kappa \dot{P}-12\beta\ddot{P}=\left[H^3P_X-6(\dot{X}_{\textrm{G}}+\dot{\Theta})H\right]_X\ .
\end{equation}
We now invoke a perturbation analysis at first order in $\alpha$, as: 
\begin{equation}
\begin{array}{ll}
\displaystyle P\simeq P^{(00)}+\kappa \left[P^{(10)}-\alpha P^{(11)}\right]\ ,
 \quad \mbox{\ }\\[8pt]
 \displaystyle D_{p, \perp}\simeq D_{p, \perp}^{(00)}+\kappa \left[D_{p, \perp}^{(10)}-\alpha D_{p, \perp}^{(11)}\right]\ ,
 \quad \mbox{\ }\\[8pt]
 \displaystyle D_{p, \parallel}\simeq D_{p, \parallel}^{(00)}+\kappa \left[D_{p, \parallel}^{(10)}-\alpha D_{p, \parallel}^{(11)}\right]\ ,
\quad \mbox{\ }\\[8pt] 
\displaystyle D_{\sigma, \parallel}\simeq D_{\sigma, \parallel}^{(00)}+\kappa \left[D_{\sigma, \parallel}^{(10)}-\alpha D_{\sigma, \parallel}^{(11)}\right]\ .
\end{array}
\label{D1}
\end{equation}
Solving Eq.~(\ref{C2}) at first order in $\alpha$, with vanishing $P^{(11)}$ at $X\rightarrow\pm \infty$, leads to:
\begin{widetext}
\begin{equation}\label{D3}
\begin{split}
& D_{p, \perp}=-\frac{3\pi}{2}\frac{\dot{\Delta}}{\Delta^{3/2}}
+\kappa\left[\frac{45\pi \ddot{\Delta}}{16 \Delta^{7/2}}-\frac{63\pi \dot{\Delta}^2}{8 \Delta^{9/2}}+\frac{3\pi \left(\dot{\Theta}-\dot{X}_{\textrm{G}}\right)^2}{8 \Delta^{7/2}}\right]-\beta\left[\frac{45\pi\dddot{\Delta}}{16\Delta^{7/2}}-\frac{567\pi\dot{\Delta}}{32\Delta^{9/2}}\left(\ddot{\Delta}-\frac{\dot{\Delta}^2}{\Delta}\right)\right.\\
&\left.+\frac{21\pi\dot{\Delta}}{32\Delta^{9/2}}\left(-6\dot{X}_{\textrm{G}}^2 -\dot{\Theta}^2+7\dot{X}_{\textrm{G}}\dot{\Theta}\right)+\frac{3\pi\ddot{\Theta}}{2\Delta^{7/2}}\left(\frac{\dot{\Theta}}{4}-\frac{7\dot{X}_{\textrm{G}}}{8}\right)  +\frac{3\pi\ddot{X}_{\textrm{G}}}{2\Delta^{7/2}}\left(\frac{3\dot{X}_{\textrm{G}}}{2}-\frac{7\dot{\Theta}}{8}\right)\right]\ ,
\end{split}
\end{equation}
\end{widetext}
\begin{widetext}
\begin{equation}\label{D4}
\begin{split}
& D_{p, \parallel}=\pi \sqrt{2\epsilon}\frac{\dot{\Theta}-\dot{X}_{\textrm{G}}}{\Delta^{1/2}}+\kappa\sqrt{\frac{\epsilon}{2}}\left[\frac{23\pi \dot{\Delta}(\dot{\Theta}-\dot{X}_{\textrm{G}})}{8 \Delta^{7/2}}+\frac{\pi(\ddot{X}_{\textrm{G}}-\ddot{\Theta})}{2 \Delta^{5/2}}\right]-\beta\sqrt{\frac{\epsilon}{2}}\left[-\frac{21\pi\dot{\Delta}^2}{4\Delta^{9/2}}\left(\dot{\Theta}-\frac{7\dot{X}_{\textrm{G}}}{4}\right)\right.\\
&\left.-\frac{9\pi\dot{X}_{\textrm{G}}}{8\Delta^{7/2}}(\dot{\Theta}-\dot{X}_{\textrm{G}})^2 + \frac{\pi\ddot{\Delta}}{16\Delta^{7/2}}(43\dot{\Theta}-73\dot{X}_{\textrm{G}}) +\frac{\pi\dot{\Delta}}{16\Delta^{7/2}}(23\ddot{\Theta}-53\ddot{X}_{\textrm{G}})+\frac{\pi}{2\Delta^{5/2}}(\dddot{X}_{\textrm{G}}-\dddot{\Theta})\right]\ ,
\end{split}
\end{equation}
\end{widetext}
\begin{widetext}
\begin{equation}\label{D5}
\begin{split}
& D_{\sigma, \parallel}=-\pi\sqrt{2\epsilon}\frac{\dot{\Theta}}{\Delta^{1/2}}+\kappa\sqrt{\frac{\epsilon}{2}}\left[\frac{\pi (\ddot{\Theta}-\ddot{X}_{\textrm{G}})}{4 \Delta^{5/2}}+\frac{\pi \dot{\Delta} \dot{X_{\textrm{G}}}}{2\Delta^{7/2}}-\frac{19\pi \dot{\Delta} \dot{\Theta}}{8\Delta^{7/2}}\right]-\beta\sqrt{\frac{\epsilon}{2}}\left[\frac{21\pi\dot{\Delta}^2}{32\Delta^{9/2}}(7\dot{\Theta}-5\dot{X}_{\textrm{G}})\right.\\
&\left.+\frac{9\pi\dot{X}_{\textrm{G}}}{16\Delta^{7/2}}(\dot{\Theta}-\dot{X}_{\textrm{G}})^2 -\frac{\pi\ddot{\Delta}}{4\Delta^{7/2}}\left(11\dot{\Theta}-\frac{29}{4}\dot{X}_{\textrm{G}}\right) -\frac{\pi\dot{\Delta}}{4\Delta^{7/2}}\left(\ddot{\Theta}-\frac{19}{4}\ddot{X}_{\textrm{G}}\right)-\frac{\pi}{4\Delta^{5/2}}(\dddot{X}_{\textrm{G}}-\dddot{\Theta})\right]\,.
\end{split}
\end{equation}
\end{widetext}

\section{Numerical integration }\label{appd}
In order to solve Eqs.~(11),~(12), and~(13) of the main text, we use the \verb|DifferentialEquations.jl| library from the open-source software for Scientific Machine Learning (SciML) collection in Julia. This choice was motivated by the strong variable coupling in our problem, which requires the use of the mass-matrix formalism. Moreover, stiffness was expected due to the third-order time derivative terms. Therefore, we employed the Rosembrock method of fifth order (\verb|Rodas5P|). It is an A-stable method which is compatible with the mass-matrix formalism and has a stable adaptative time step. However, there were still some need for fine tuning the solver's tolerances, mainly around \verb|1e-8| for both absolute and relative tolerances. Finally, we noted that the solver fails to simulate the sedimentation behaviour at long times. This is likely due to $\Delta$ flipping sign when approaching zero, because of the limitations on the smallest step size available.